\documentstyle[amssymb,twocolumn,aps,prl,epsfig]{revtex}
\twocolumn

\begin{document}
\title{Inhibition of electromagnetically induced absorption due to excited state
decoherence in Rb vapor. }
\author{H. Failache, P. Valente, G. Ban\thanks{%
Permanent address: Laboratoire de Physique Corpusculaire, 14050 Caen, France.%
}, V. Lorent\thanks{%
Permanent address: Laboratoire de Physique des Lasers, 93430 Villetaneuse,
France.} and A. Lezama\thanks{%
E-mail: alezama@fing.edu.uy}}
\address{Instituto de F\'{i}sica, Facultad de Ingenier\'{i}a. Casilla de correo 30. \\
11000, Montevideo, Uruguay.}
\date{\today }
\maketitle

\begin{abstract}
The explanation presented in [Taichenachev et al, Phys. Rev. A {\bf 61},
011802 (2000)] according to which the electromagnetically induced absorption
(EIA) resonances observed in degenerate two level systems are due to
coherence transfer from the excited to the ground state is experimentally
tested in a Hanle type experiment observing the parametric resonance on the $%
D1$ line of $^{87}$Rb. While EIA\ occurs in the $F=1\rightarrow F^{\prime
}=2 $ transition in a cell containing only $Rb$ vapor, collisions with a
buffer gas ($30\ torr$ of $Ne$) cause the sign reversal of this resonance as
a consequence of collisional decoherence of the excited state. A theoretical
model in good qualitative agreement with the experimental results is
presented.
\end{abstract}

\pacs{42.50.Gy, 32.80.Bx, 03.65.Yz, 34.90.+q.}

\preprint{}

\section{Introduction.}

There has been considerable interest in recent years for the fascinating
properties of coherently prepared atomic media \cite{SCULLYBOOK}. Among the
most studied coherent effects is the phenomenon of electromagnetically
induced transparency (EIT) \cite{HARRIS97}. EIT has generally been modelled
and experimentally studied in three-level $\Lambda $ systems with two long
lived lower (ground) states and a rapidly decaying upper (excited) state.
The two electromagnetic fields (pump and probe) separately couple each of
the two arms of the $\Lambda $ system. A distinctive feature of the EIT
resonance is its narrow linewidth which corresponds to the coherence decay
rate of the ground state doublet. The occurrence of the EIT resonance is
directly linked to the existence of a dark state (DS): i.e. a linear
combination of the two ground states uncoupled to the excited state. EIT is
a consequence of the system being pumped into the DS.

EIT may also be observed in multi-level systems as those involving the
Zeeman substates of two atomic levels with angular momentum degeneracy,
hereafter called degenerate two-level systems (DTLS). In such case too, the
observation of EIT is a direct consequence of the existence of a DS within
the lower atomic level when $F_g\geq F_e$ ($F_g$ and $F_e$ are the angular
momenta of the ground and excited state respectively).

The pump-probe spectroscopy of DTLS with $0<F_{g}<F_{e}$ also presents
resonances in the probe transmission when the Raman resonance condition
between ground state Zeeman sublevels is fulfilled. However, in this case,
the resonances correspond to an increase of the probe absorption and have
consequently being designated as electromagnetically induced absorption
(EIA) \cite{AKULSHIN98,LEZAMA99,LEZAMA00}. As in the case of EIT resonances,
the EIA linewidth is given by the coherence decay rate of the ground level.
Unlike EIT, the EIA resonances cannot be associated to the existence of a DS
in the ground state. Also, there is no simple connection between the EIA
resonances and the existence of a DS in the {\em excited} state since such
state has a lifetime limited by spontaneous emission and cannot account for
the narrow spectral features observed.

An explanation for the EIA resonance was provided by Taichenachev et al \cite
{TAICHENACHEV00} analyzing a four level system in a $N$ configuration. They
showed theoretically that in this system, which is the simplest to present
EIA, the enhanced absorption is due to the transfer, via spontaneous
emission, of the coherence created by the exciting fields within the two
upper levels. Although the $N$ configuration does not correspond to the
configurations actually explored in realistic DTLS, the argument presented
in \cite{TAICHENACHEV00} can be extended to such systems \cite{VALENTE02}.

The purpose of this paper is to provide experimental evidence in support of
the argument presented in \cite{TAICHENACHEV00} by demonstrating that EIA
resonances are suppressed (and even reversed) if the coherence of the
excited state is significantly destroyed by collisions before the occurrence
of the spontaneous emission decay. A simple theoretical model in good
agreement with the observations is also presented.

A convenient experimental scheme for the observation of coherence resonances
is the Hanle type setup that uses a unique optical beam with linear
polarization in near resonance with an atomic transition. In this scheme,
the two opposite circular polarization components of the light can be
considered as the pump and probe fields. The light beam is sent through an
atomic sample where the Raman resonance condition is tuned, via the Zeeman
effect, by a magnetic field along the light propagation axis. The intensity
of the transmitted light is monitored. Hanle/EIT/EIA resonances on the $D1$
lines of alkaline vapors were recently studied by several groups \cite
{RENZONI97,RENZONI98,RENZONI99,DANCHEVA00,RENZONI01PRA,ALZETTA01} (see inset
in Fig. \ref{isotope} for an energy level scheme of the $^{87}$Rb $D1$
line). Dancheva et al \cite{DANCHEVA00} were the first to observe a
Hanle/EIA resonance on a $F_{g}\rightarrow F_{e}=F_{g}+1$ transition of the $%
D1$ line. They pointed out the fact that the EIA resonance can be observed
in spite of the transition being open (radiative decay can occur from the
excited state to either ground state hyperfine levels). The open character
of the transition is responsible for the smallness of the Hanle/EIA
resonance as compared with the Hanle/EIT resonances occurring when $%
F_{e}=F_{g},\ F_{g}-1$.

\begin{figure}[tbp]
\begin{center}
\mbox{\epsfig{file=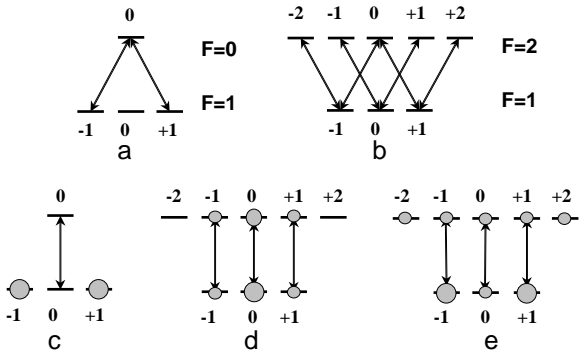,width=3.5in}}
\end{center}
\caption{Examples of energy level configurations in DTLS with a linear
optical field polarization. $a$ and $b$: quantization axis perpendicular to
the optical polarization. $c-d$: quantization axis parallel to the optical
polarization. The circles represent the sublevel population. $d$: no
collisional relaxation of the excited state. $e$: collisionaly thermalized
excited state.}
\label{levelschemes}
\end{figure}

The choice of the Hanle experimental scheme for coherent spectroscopy is not
only motivated by its simplicity. It also allows a special insight on the
connection between coherence resonance and optical pumping. Consider the
situation depicted in Fig. \ref{levelschemes}. Figs. \ref{levelschemes}a)
and \ref{levelschemes}b) correspond to a $F_{g}=1\rightarrow F_{e}=0$ and $%
F_{g}=1\rightarrow F_{e}=2$ transitions respectively. In both cases, the
quantization axis has been chosen parallel to the direction of the light
propagation which coincides with the orientation of the magnetic field. The
Raman resonance condition is achieved for the two components of the optical
field at zero magnetic field and destroyed for nonzero magnetic field. In
Fig.\ref{levelschemes}a) the optical field interact with a $\Lambda $ system
for which EIT is known to occur as a consequence of the existence of a DS
formed by the anti-symmetric combination of the ground state Zeeman
sublevels $\left| -1\right\rangle $ and $\left| +1\right\rangle $. The
situation in Fig. \ref{levelschemes}b) corresponds to EIA which results,
according to \cite{TAICHENACHEV00}, from the transfer of coherence from the
upper to the lower level via spontaneous emission. Figs. \ref{levelschemes}%
c) and \ref{levelschemes}d) represent the same physical situation than in a)
and b) respectively with the quantization axis taken along the direction of
the linear polarization of the light. In Fig. \ref{levelschemes}c) the
transparency observed for zero magnetic field is the consequence of the
optical pumping of the system into the $\left| -1\right\rangle $ and $\left|
+1\right\rangle $ sublevels which are not coupled to the excited state. In
Fig. \ref{levelschemes}d) the absorption increase for $B=0$ is a consequence
of the redistribution of the population of the ground state Zeeman sublevels
(alignment) via optical pumping. The atomic population accumulates
preferentially in the $\left| 0\right\rangle $ sublevel which has the
largest coupling with the excited state. Notice that with this choice of the
quantization axis, no coherence is built among the excited state Zeeman
sublevels and thus cannot be transferred to the ground state. In the basis
used in c) and d) the Hanle/EIT and Hanle/EIA resonances appear as a
consequence of incoherent optical pumping.

Since the choice of the quantization axis is arbitrary, the two frames
considered in the previous discussion can conveniently be considered for the
analysis of the Hanle/EIA resonances under the effect of collisions
affecting the excited state. In the scheme of Fig. \ref{levelschemes}b)
which can be considered as an extension of the simple $N$ system analyzed by
Taichenachev et al \cite{TAICHENACHEV00}, EIA is a consequence of the
spontaneous transfer of Zeeman coherence from the excited to the ground
state. In consequence, the EIA resonance should disappear if the excited
state coherence is destroyed by collisions in a time comparable or shorter
than the excited state lifetime. In the basis corresponding to Fig. \ref
{levelschemes}d) EIA can be prevented if the collisions are responsible for
a significant equalization (thermalization) of the excited state sublevel
populations before the occurrence of spontaneous emission (Fig. \ref
{levelschemes}e). If the excited state is completely thermalized then the
total spontaneous emission decay into a given ground state Zeeman sublevel
is the same for all ground state sublevels. In consequence, in the steady
state, the population is preferentially accumulated in sublevels $\left| \pm
1\right\rangle $ resulting in an increased transparency.

To demonstrate the influence of the excited state coherence on EIA we have
observed the Hanle/EIT/EIA resonances corresponding to the four transitions
of the $D1$ line of $^{87}$Rb both in a vapor cell containing only Rb vapor
(negligible collisions) and in a cell containing $30\ torr$ of Ne as buffer
gas.

It is well known that room temperature collisions with light noble gas atoms
(He, Ne, Ar) produce different effects on the $^2S_{1/2}$ ground state than
in the $^2P_{1/2}$ or $^2P_{3/2}$ excited states of several alkaline vapors 
\cite{HAPPER72}. The collisions of a noble gas atom with an alkaline atoms
in its ground state has little effect on its electronic and nuclear spin. As
a consequence the alkaline atom can experience a very large number of
collisions while preserving the ground state coherence. However, the atomic
motion is affected by collisions becoming diffusive. This results in a
longer interaction time before the atoms can leave the light beam or reach
the cell walls. The increased interaction time allows the observation of
coherence resonances with time-of-flight limited linewidth of only a few
tenth of $Hz$ \cite{BRANDT97,ERHARD00,ERHARD01}.

The situation is rather different for collisions between noble gas atoms and
alkaline atoms in the $^{2}P_{1/2}$ state\cite{HAPPER72}. At room
temperature and for the buffer gas density corresponding to the experiments
described below, the collisions are sufficiently energetic and frequent to
produce a considerable broadening of the homogeneous width of the optical
transitions (which remains nevertheless smaller than the Doppler width and
the $^{2}P_{1/2}$ hyperfine splitting). Also under such conditions virtual
transitions to the neighboring levels (mainly $^{2}P_{3/2}$) occur \cite
{GALLAGHER67,BULOS71} resulting in the non preservation of the magnetic
quantum number ($m_{J}$) of the electronic angular momentum during
collisions. As a consequence, a significant thermalization of the excited
state density matrix take place in a time shorter than the excited state
lifetime. The cross section for excited state collisional $m_{J}$ mixing was
measured by several authors \cite
{GALLAGHER67,BULOS71,KEDZIERSKI94,ROTONDARO98}. Using the figures in \cite
{BULOS71} and the Ne density corresponding to the experiments the
collisional decoherence rate of the excited state $\gamma _{coll}$ can be
estimated as $\gamma _{coll}\approx 4\Gamma $ where $\Gamma $ is the excited
state spontaneous emission decay rate.

\section{Experiment}

\begin{figure}[tbp]
\begin{center}
\mbox{\epsfig{file=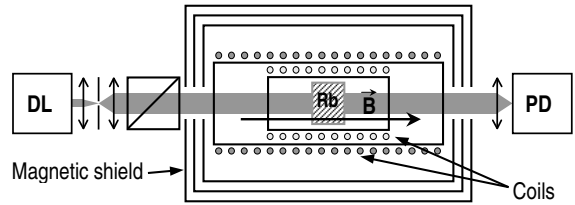,width=3.5in}}
\end{center}
\caption{Experimental setup. DL: diode laser. PD: photodiode.}
\label{setup}
\end{figure}

The experimental setup is shown in Fig. \ref{setup}. We have used two
cylindrical vapor cells of the same dimensions (diameter: $2.5\ cm$, length: 
$5\ cm$) provided by the same manufacturer and filled with natural rubidium
vapor. One of the cells contains $30\ torr$ of Ne as buffer gas. The cell
under consideration was placed (at room temperature) inside a magnetic
shield formed by three coaxial cylindrical $\mu $-metal layers. After
degaussing of the $\mu $-metal shield the total inhomogeneity of the
residual magnetic field at the vapor cell was less than $10\ \mu G$. A
coaxial solenoid, internal to the magnetic shield was used to scan the
magnetic field at the atomic sample.

An extended cavity diode laser ($\sim 1\ MHz$ linewidth) was used for the
atomic excitation. The laser frequency was monitored and stabilized on
specific hyperfine transitions with the help of an auxiliary saturated
absorption setup. The total laser power at the sample was approximately $%
0.1\ mW$. The light transmitted through the vapor cell was monitored with a
photodiode ($100\ kHz$ bandwidth).

\begin{figure}[tbp]
\begin{center}
\mbox{\epsfig{file=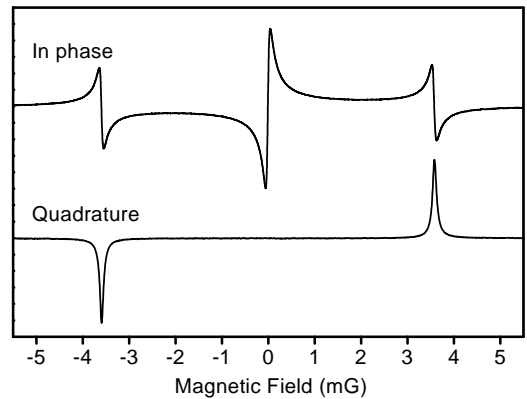,width=3.5in}}
\end{center}
\caption{Experimental parametric resonance signal for the $%
F_{g}=2\rightarrow F_{e}=1$ transition of $^{87}Rb$ in the cell containing
buffer gas. Modulation frequency $f=5\ kHz$. The two traces are on the same
vertical scale (shifted for clarity). }
\label{expbuffbroad}
\end{figure}

In order to enhance sensitivity the coherence resonances were detected
through a parametric resonance technique \cite{DUPONT70}. In addition to the
slowly varying DC magnetic field, a small AC component of the magnetic field
(oriented along the light propagation axis) was introduced with a secondary
coil driven at frequency $f$. A lock-in amplifier detected the phase and
quadrature components of the photodiode current oscillating at frequency $f$%
. Two different values of $f$ were used in the measurements. For the cell
without buffer gas we used $f=75\ kHz$ and for the cell containing Ne $f=5\
kHz$ was used. In both cases $f$ was chosen to exceed the width of the
corresponding Hanle resonances. A typical recording of the lock-in output
signal as a function of the solenoid DC current is shown in Fig. \ref
{expbuffbroad}. The central structure of the spectrum corresponds to $B=0$.
The spectrum sidebands occurring for $2\Delta E_{Z}=hf$ ($\Delta E_{Z}$ is
the ground state Zeeman energy shift) allow a precise calibration of the
magnetic field inside the solenoid.

\begin{figure}[tbp]
\begin{center}
\mbox{\epsfig{file=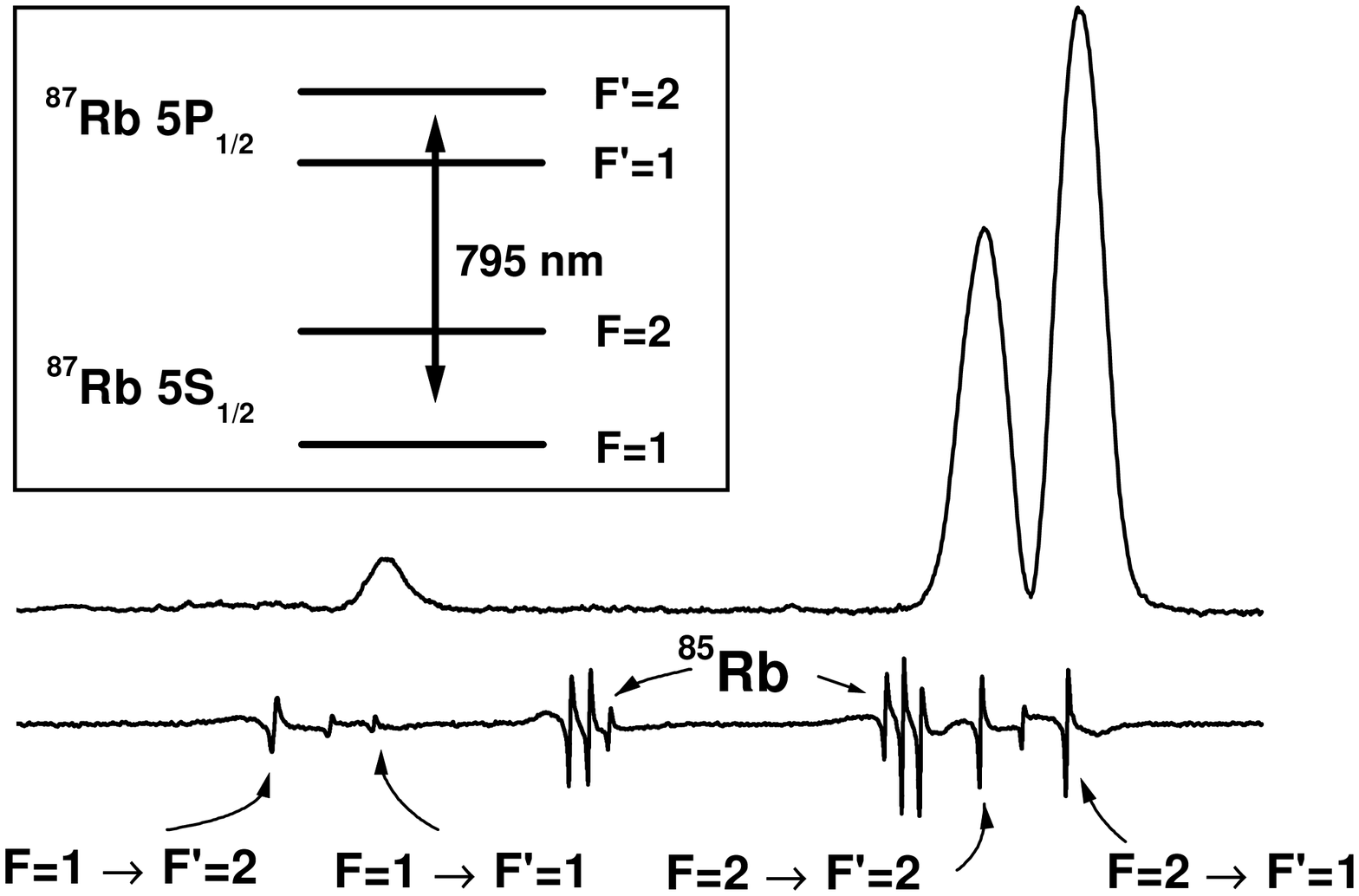,width=3.5in}}
\end{center}
\caption{Quadrature parametric resonance signal as a function of the laser
frequency (upper trace). The DC magnetic field is kept fixed at the value
corresponding to the positive sideband on Fig. \ref{expbuffbroad}. Lower
trace: reference saturated absorption signal. Inset: level scheme for the $D1
$ line transitions of $^{87}$Rb.}
\label{isotope}
\end{figure}

In the cell containing the buffer gas \ where the resonances are well
resolved (Fig. \ref{expbuffbroad}) the parametric resonance scheme used in
the experiment allows for isotope selective spectroscopy. For this, the DC
magnetic field is kept fixed at the value corresponding to the maximum of
one of the (Lorentzian) sidebands resonances observed on the quadrature
signal and the quadrature output of the lock-in amplifier is monitored while
the frequency of the laser is scanned on the $D1$ line. Since the resonance
condition $2\Delta E_{Z}\equiv 2g\mu _{B}B=hf$ depends on the specific
isotope through the gyromagnetic factor $g$ ($\mu _{B}$ is the Bohr
magneton), only one isotope contributes to the observed spectrum. An example
of such spectrum is shown in Fig. \ref{isotope} where only the $D1$
transitions of $^{87}$Rb appear in spite of the cell being filled with
natural Rb ($72\%$ of $^{85}$Rb). The gyromagnetic factors of the two
hyperfine levels of the ground state of $^{87}$Rb differ in absolute value
by less than $0.3\%$. Consequently all four hyperfine transitions can be
observed in the same spectrum. Notice that the hyperfine structure is well
resolved in the spectrum of Fig.\ref{isotope}. This is a clear indication
that in spite of the rather strong collisional regime, the atomic level
structure resulting from the hyperfine coupling is preserved and that the
total angular momentum $F_{e}$ remains a good quantum number. The relative
weight of the coherence resonances in the four $D1$ transitions is
appreciated in Fig. \ref{isotope}. The peak corresponding to the $%
F_{g}=1\rightarrow F_{e}=2$ transition is smaller than that of the $%
F_{g}=1\rightarrow F_{e}=1$ transition by a factor $25$ and is not visible
on the scale of Fig. \ref{isotope}. It is worth mentioning that the small $%
F_{g}=1\rightarrow F_{e}=2$ signal is nevertheless $100$ times larger than
the residual signal due to the wings of the $F_{g}=1\rightarrow F_{e}=1$
peak estimated from a Gaussian fit.

\begin{figure}[tbp]
\begin{center}
\mbox{\epsfig{file=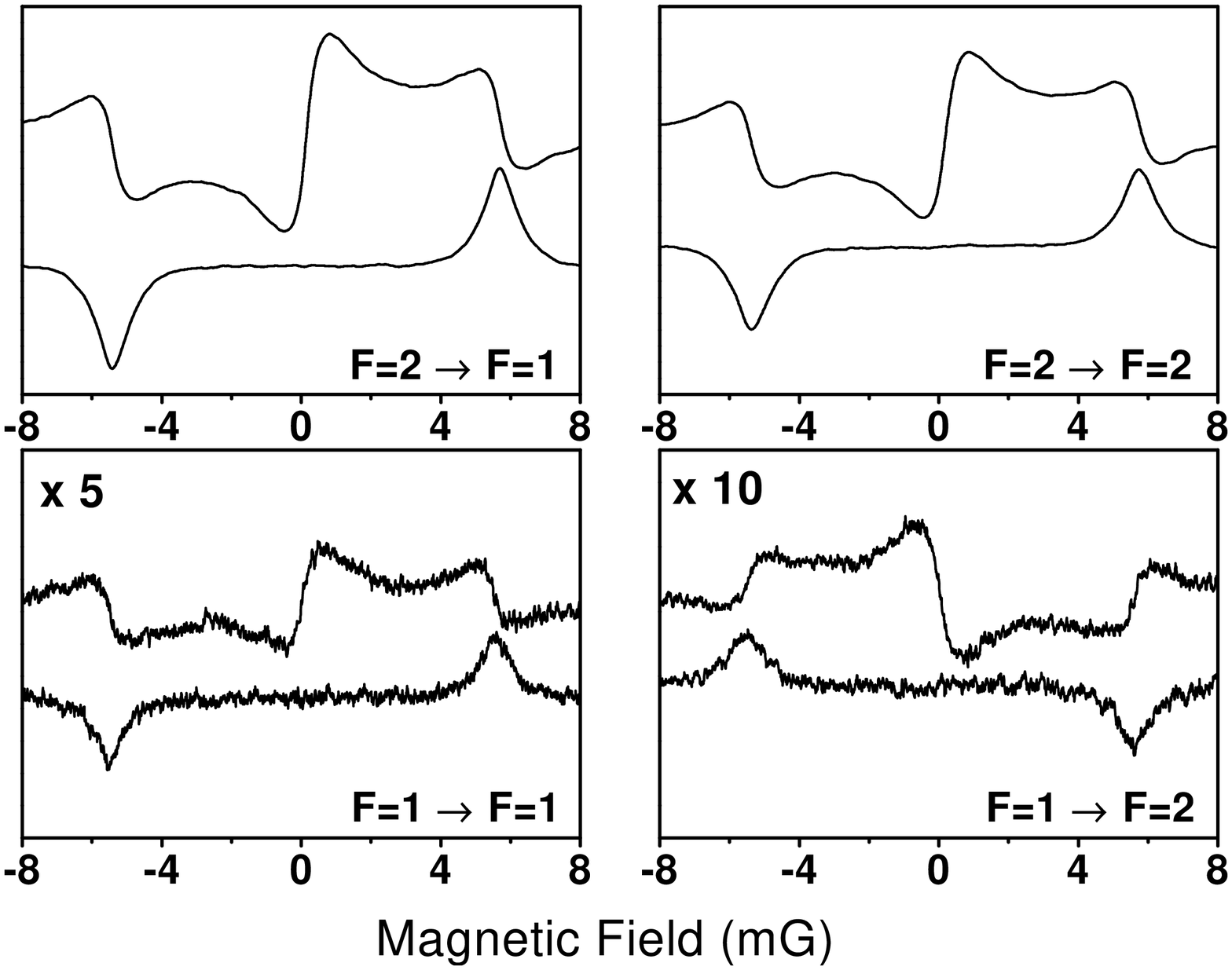,width=3.5in}}
\end{center}
\caption{In phase (upper trace) and quadrature (lower trace) parametric
resonance signal for the four hyperfine transitions of the $D1$ line of $%
^{87}$Rb in the cell without buffer gas. Comparable vertical scales are used
for all curves.}
\label{expnobuffer}
\end{figure}

\begin{figure}[tbp]
\begin{center}
\mbox{\epsfig{file=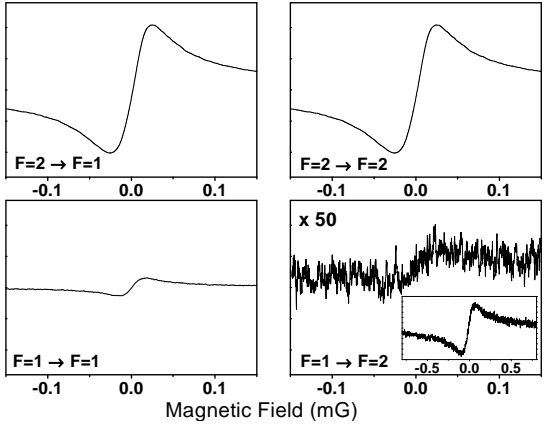,width=3.5in}}
\end{center}
\caption{Central structure of the in phase parametric resonance signal for
the four hyperfine transitions of the $D1$ line of $^{87}$Rb in the cell
containing $30\ torr$ of Ne. Comparable vertical scales are used for all
curves. The inset shown with the $F=1\rightarrow F=2$ \ transition was
recorded with increased Rb density.}
\label{expbuffer}
\end{figure}

Figs. \ref{expnobuffer} and \ref{expbuffer} show the observed signals with
the laser in resonance with the four $D1$ transitions in the cell without
buffer gas and in the cell containing Ne respectively using the same optical
power. The beam diameter was $10\ mm$ and $5\ mm$ respectively. Only the
central resonance around $B=0$ for the in phase signal is shown in Fig. \ref
{expbuffer}. Without buffer gas the observed resonances have a width which
is determined by the time of flight through the optical beam. Notice the
sign reversal of the $F_{g}=1\rightarrow F_{e}=2$ transition (EIA)\ with
respect to the three other resonances (EIT). In the cell with buffer gas
(Fig. \ref{expbuffer}), the observed resonances are much narrower as
expected from the increase in interaction time due to the diffusive atomic
motion. However, under the present experimental conditions the observed
linewidth is limited by residual magnetic field inhomogeneities. A sign
change is clearly observed for the $F_{g}=1\rightarrow F_{e}=2$ transition
with respect to the cell without Ne demonstrating the quenching of the EIA
resonance and its reversal into EIT.

\section{Model}

We present in this section a simple theoretical model containing the
essential ingredients for the analysis of Hanle/EIT/EIA coherence resonances
in DTLS. The model takes explicitly into account the Zeeman degeneracy of
the ground and excited states. It considers a unique electric dipole allowed
atomic transition in a homogeneous sample of atoms at rest. With respect to
the conditions of the experiment, the model contains several
simplifications. The effect of the atomic motion, the spatial distribution
of the light field, the light propagation in the sample and the influence of
neighboring transitions are not considered. Also the effect of collisions is
introduced in an quite simplified way through a single collisional
relaxation rate as discussed below.

We consider a two-level transition between a ground state $g$ of angular
momentum $F_{g}$ and an excited state $e$ of angular momentum $F_{e}$ with
energy separation $\hbar \omega _{0}$. The atoms are illuminated by an
optical field of amplitude $E$ and frequency $\omega $ linearly polarized
along the unit vector ${\bf e}$ and submitted to a magnetic field $B$
perpendicular to ${\bf e}$. The atoms are submitted to collisions with the
buffer gas. It is assumed that the collisions result in dephasing of the
atomic optical dipole and can cause real transitions between excited state
Zeeman sublevels.

In the frame rotating with the optical field and with the usual rotating
wave approximation, the Liouville equation for the density matrix $\sigma $
of the system is: 
\begin{eqnarray}
\dot{\sigma} &=&-\frac{i}{\hbar }\left[ \hbar \Delta P_{e}+H_{B}+V,\sigma %
\right] -\frac{\left( \Gamma +\gamma _{coll}\right) }{2}\left\{ P_{e},\sigma
\right\}  \nonumber \\
&&+b\Gamma \sum_{q=-1,0,1}Q_{ge}^{q}\sigma Q_{eg}^{q}-\gamma \sigma +\gamma
\sigma _{0}+\gamma _{coll}\bar{\sigma}  \label{Liouville}
\end{eqnarray}
where $\Delta =\omega _{0}-\omega $, $H_{B}=-\mu _{B}F_{z}\left(
g_{g}P_{g}+g_{e}P_{e}\right) B\equiv -M_{z}B\ $ is the Zeeman Hamiltonian ($%
P_{g}$ and $P_{e}$ are projectors on the ground and excited subspaces
respectively, $g_{g}$ and $g_{e}$ are the gyromagnetic factors of the ground
and excited states respectively, $\mu _{B}$ is the Bohr magneton and $\hbar
F_{z}$ the projection of the total angular momentum along the direction of
the magnetic field).

The atom field interaction is given by: 
\begin{equation}
V=\frac{\hbar \Omega }{2}{\bf e}\cdot {\bf Q}  \label{atomfield}
\end{equation}
${\bf Q}$ is a dimensionless vectorial operator related to the electric
dipole operator ${\bf D}$ through: 
\begin{equation}
{\bf D}\equiv {\bf Q}\left\langle F_{g}\left\| {\bf D}\right\|
F_{e}\right\rangle  \label{operatorq}
\end{equation}
$\Omega \equiv \frac{\left\langle F_{g}\left\| {\bf D}\right\|
F_{e}\right\rangle E}{\hbar }$ is the reduced Rabi frequency of the optical
field ($\left\langle F_{g}\left\| {\bf D}\right\| F_{e}\right\rangle $ is
the reduced matrix element of the electric dipole operator for the
considered transition). $Q_{ge}^{q}=Q_{eg}^{q\dagger }$ are the spherical
components of the operator $P_{g}{\bf Q}P_{e}$. $b$ is a branching ratio
coefficient ($0\leq b\leq 1$). $\Gamma $ is the spontaneous emission decay
rate. $\gamma $ and $\gamma _{coll}$ are relaxation rates associated to
transit time and excited state collisions respectively. $\sigma _{0}=\frac{%
P_{g}}{2F_{g}+1}$ corresponds to an isotropic density matrix for the ground
level with unit total population and $\bar{\sigma}=\frac{P_{e}}{2F_{e}+1}%
Tr\left( P_{e}\sigma \right) $ is an (incoherent) isotropic density matrix
for the excited state with the same excited state population than $\sigma $.

The first term on the r.h.s. of Eq. \ref{Liouville} describes the
Hamiltonian evolution of the atom in the presence of the optical and
magnetic field. The second term account for the relaxation of the excited
state and the optical coherences. In addition to spontaneous emission we
consider the relaxation due to collisions with the buffer gas atoms. For
simplicity, it is assumed that the effect of buffer gas collisions on
optical and Zeeman coherences can be described by the single relaxation rate 
$\gamma _{coll}$. The third term on the r.h.s. of Eq. \ref{Liouville}
describes the spontaneous emission transfer of population and coherence from
the excited state to the ground state. For open transitions, the branching
ratio coefficient $b$ accounts for the atomic loss due to radiative
transitions to external levels ($b=1$ corresponds to closed transitions).
The fourth term on the r.h.s. of Eq. \ref{Liouville} accounts, in the usual
phenomenological way, for the finite interaction time. Although this
relaxation term concerns both the ground and the excited level, notice that $%
\gamma $ is the only relaxation rate acting on the ground state. The escape
of atoms from the interaction region at rate $\gamma $ is compensated at
steady state by the arrival of ''fresh'' atoms isotropically distributed in
the ground state ($\gamma \sigma _{0}$ in Eq. \ref{Liouville}). The last
term on the r.h.s. of Eq. \ref{Liouville} is an effective isotropical
repumping term introduced to compensate the effect of the collisions
(included in the second term of the r.h.s. of Eq. \ref{Liouville}) on the
total excited state population.

The steady state solution of Eq. \ref{Liouville} can be easily obtained
numerically \cite{LEZAMA99,LEZAMA00}. The solution of Eq. \ref{Liouville}
for conditions corresponding to the parametric resonance scheme where the
magnetic field along the light propagation axis is the sum of a DC and a
small sine modulated component is presented in the appendix. The calculation
of the light absorption signals in phase and in quadrature with respect to
the magnetic field modulation are also derived in the appendix.

\begin{figure}[tbp]
\begin{center}
\mbox{\epsfig{file=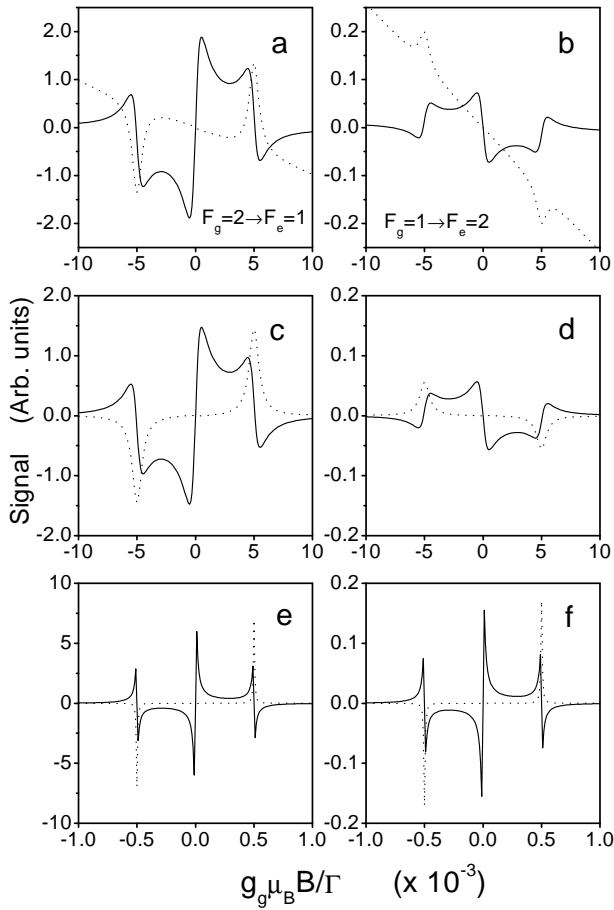,width=3.5in}}
\end{center}
\caption{Calculated signal for the $F_{g}=2\rightarrow F_{e}=1$ (left
column) and $F_{g}=1\rightarrow F_{e}=2$ (right column) transitions ($\Omega
=0.01\Gamma $). Solid: in phase signal. Dotted: quadrature signal. $a,b,c,d$%
: $\protect\gamma _{coll}=0$, $\protect\gamma =10^{-3}\Gamma $, $2\protect\pi
f=10^{-2}\Gamma $. $e,f$: $\protect\gamma _{coll}=4\Gamma $, $\protect\gamma %
=10^{-5}\Gamma $, $2\protect\pi f=10^{-3}\Gamma $. In $c$ and $d$ the signal
was averaged over $\Delta $ in the range $-5\Gamma <\Delta <5\Gamma $. The
vertical scales are independent for each row but comparable within a row. }
\label{teo}
\end{figure}

The prediction of the model for the transitions of the $D1$ line of $\ ^{87}$%
Rb are presented in Fig. \ref{teo}. Only the $F_{g}=2\rightarrow F_{e}=1$
and $F_{g}=1\rightarrow F_{e}=2$ transitions are shown. The parameters of
the calculation were chosen to correspond approximately to the experimental
conditions. The same light intensity is assumed in all figures. We used $%
\Omega /\Gamma =0.01$ (the two transitions shown have the same strength).
The branching ratio was taken as $b=0.83$ and $b=0.5$ respectively
(evaluated through standard angular momentum calculus) and the signal
corresponding to each transition was weighted proportionally to the thermal
occupation number of the lower level (coefficients $5/8$ and $3/8$
respectively). The plots a-d in Fig. \ref{teo} correspond to the absence of
collisions. In these plots the transit time relaxation rate was taken as $%
\gamma =10^{-3}\Gamma $. While plots a,b,e,f correspond to the direct output
of our model for $\Delta =0$, plots c and d correspond to the signal
averaged over $\Delta $ in the range $-5\Gamma <\Delta <5\Gamma $ to include
the contribution of different velocity classes to the signal ($\Delta =-kv$, 
$k$ is the wavevector and $v$ the velocity). When different from zero, the
collisional rate was taken as $\gamma _{coll}=4\Gamma $ a value estimated
from the data in \cite{BULOS71}. The precise value of $\gamma _{coll}$ has
small influence on the lineshape of the signal provided that it is taken
sufficiently larger than $\Gamma $. Notice the sign reversal on the
calculated signal corresponding to the $F_{g}=1\rightarrow F_{e}=2$
transition when $\gamma _{coll}\neq 0$ with respect to the case in which $%
\gamma _{coll}=0.$

\section{Discussion.}

In spite of the simplicity of the model, the main features of the
experimental observations are well reproduced in the calculated spectra
(similar agreement is obtained for the two hyperfine transitions not shown).
The variation in the amplitude of the signal between the two hyperfine
transitions is essentially the consequence of the different values of the
branching ratio $b$. In the plots a and b of Fig. \ref{teo} a non resonant
contribution is clearly visible giving rise to the observed slope in the
quadrature signal. Such non resonant contribution is negligible on the scale
of plots e and f. We attribute this difference to the fact that the non
resonant contribution originates in the linear response of the atomic sample
while the resonances are due to its nonlinear response. In an open two level
system, the saturation intensity $\Omega _{S}^{2}$ is of the order of $%
\gamma \Gamma $. With the values of $\gamma $ used in the two cases
considered above, we have $\Omega <\Omega _{S}$ for the pure Rb vapor (plots
a - d) and $\Omega >\Omega _{S}$ for plots e and f. This explains the
different relative weight of the linear and nonlinear contributions in the
two cases. Since the coherence resonance condition is essentially
Doppler-free, only when $\Omega <\Omega _{S}$ the integration over different
detunings $\Delta $ (integration over velocity classes ) result in a
significant change in the lineshape. After integration, the nonresonant
background is cancelled (assuming that the laser is centered on the Doppler
profile) in agreement with the experimental observation.

Several simplifications were assumed in the theoretical model. The
approximation consisting in considering an homogeneous sample of atoms at
rest have already been addressed; when necessary this simplification can be
abandoned carrying on the integration over velocity classes as described
above. Another important approximation is the simplified treatment of the
collisional process described through a unique scalar rate. Also, it is
assumed in the model that the atom remain confined to the excited state
level during the collisional process. However, for the Ne density used in
the experiment, inelastic collisions have a non-negligible probability and a
fraction of the excited atoms may be collisionaly transferred to a different
excited hyperfine level or even to the $5P_{3/2}$ manifold \cite{BULOS71}.
In any case, such inelastic processes are not expected to preserve the
Zeeman coherence of the excited levels. The overall consequence of the
inelastic collision processes, followed by spontaneous emission, is to
provide a second (indirect) path for the return of the atoms to the ground
state. With the assumption that this additional decay channel is isotropic
its contribution can be effectively included in the collisional rate $\gamma
_{coll}$.

\section{Conclusions.}

We have studied Hanle/EIT and Hanle/EIA\ resonances on the $D1$ line of $\
^{87}$Rb vapor both in absence and in presence of a buffer gas. The
coherence resonances were studied through a parametric resonance technique.
We observed that the Hanle/EIA\ resonance, occurring for the $%
F_{g}=1\rightarrow F_{e}=2$ transition, change sign in the presence of a
buffer gas as a consequence of collisions in the excited state. A
theoretical model was presented that allows the numerical calculation of the
parametric resonance signals. In spite of its simplicity, the model
reproduces the essential features of the experimental results on the
assumption that the collision of the Rb atoms with the buffer gas result in
the isotropic decoherence of the excited state. The overall agreement
between the experimental results and the simple model clearly suggest that
the preservation of the excited state coherence during the excited state
lifetime and its transfer to the ground state are, as suggested in \cite
{TAICHENACHEV00}, key ingredients for the occurrence of EIA.

\section{Acknowledgments.}

The authors are thankful to D. Bloch for fruitful discussions. This work was
supported by the Uruguayan agencies:\ CONICYT, CSIC and PEDECIBA and by ECOS
(France).

\section{Appendix.}

In this appendix we describe the calculation of the parametric resonance
signals.

We consider the evolution of the atom-optical field system under the
influence of a longitudinal magnetic field of the form: 
\begin{equation}
B=B_{0}+B_{1}\cos \left( \delta t\right)  \label{modulatedb}
\end{equation}
We seek a solution for the density matrix on the form: $\sigma \left(
t\right) \simeq \sigma ^{0}+\sigma ^{1}\left( t\right) $ where $\sigma ^{0}$
is constant and $\sigma ^{1}\left( t\right) $ is a time dependent correction
to first order in $B_{1}$. Substituting in Eq. \ref{Liouville}, we have: 
\begin{eqnarray}
0 &=&-\frac{i}{\hbar }\left[ \hbar \Delta P_{e}-M_{z}B_{0}+V,\sigma ^{0}%
\right] -\frac{\left( \Gamma +\gamma _{coll}\right) }{2}\left\{ P_{e},\sigma
^{0}\right\}  \nonumber \\
&&+b\Gamma \sum_{q=-1,0,1}Q_{ge}^{q}\sigma ^{0}Q_{eg}^{q}-\gamma \sigma
^{0}+\gamma \sigma _{0}+\gamma _{coll}\overline{\sigma ^{0}}
\label{orderzero}
\end{eqnarray}
\begin{eqnarray}
\dot{\sigma}^{1} &=&-\frac{i}{\hbar }\left[ \hbar \Delta
P_{e}-M_{z}B_{0}+V,\sigma ^{1}\right] +\frac{i}{\hbar }\left[
M_{z}B_{1},\sigma ^{0}\right] \cos \left( \delta t\right)  \nonumber \\
&&-\frac{\left( \Gamma +\gamma _{coll}\right) }{2}\left\{ P_{e},\sigma
^{1}\right\} +\gamma _{coll}\overline{\sigma ^{1}}  \label{orderone} \\
&&+b\Gamma \sum_{q=-1,0,1}Q_{ge}^{q}\sigma ^{1}Q_{eg}^{q}-\gamma \sigma ^{1}
\nonumber
\end{eqnarray}

If one identifies the matrix elements of $\sigma ^0$ and $\sigma ^1$ with
the components of column vectors $Y^0$ and $Y^1$, then Eqs. \ref{orderzero}
and \ref{orderone} can be rewritten as: 
\begin{equation}
{\cal M}Y^0=P  \label{Liouvorder0}
\end{equation}
\begin{equation}
\dot{Y}^1={\cal M}Y^1+A\cos \left( \delta t\right)  \label{Liouvorder1}
\end{equation}
where ${\cal M}$ is a matrix with time independent coefficients and $P$ and $%
A$ are two column vectors associated to $\gamma \sigma _0$ and $\frac i\hbar %
\left[ M_zB_1,\sigma ^0\right] $ \ respectively.

We are interested in the solution of Eq. \ref{Liouvorder1} of the form: 
\begin{equation}
Y^{1}\left( t\right) =\alpha \cos \left( \delta t\right) +\beta \sin \left(
\delta t\right)  \label{yform}
\end{equation}
after substitution in Eq. \ref{Liouvorder1} one gets: 
\begin{equation}
\alpha ={\cal M}\left( \delta ^{2}{\Bbb I}+{\cal M}^{2}\right) ^{-1}A
\label{alfa}
\end{equation}
\[
\beta =\delta \left( \delta ^{2}{\Bbb I}+{\cal M}^{2}\right) ^{-1}A 
\]
where ${\Bbb I}$ is the identity matrix.

After numerical evaluation of vectors $\alpha $ and $\beta $ one can
retrieve the corresponding density matrices $\sigma _{\alpha }$ and $\sigma
_{\beta }$ and the corresponding light absorption coefficients respectively
in phase and in quadrature with the magnetic field modulation: 
\begin{equation}
\lambda _{i}\varpropto 
\mathop{\rm Im}%
\left[ {\bf e\cdot }Tr\left( \sigma _{i}{\bf D}\right) \right] \;\ \left(
i=\alpha ,\ \beta \right)   \label{phaseandquad}
\end{equation}


\end{document}